\documentclass[aps,pra,preprint,showpacs,superscriptaddress,amsfonts,floatfix]{revtex4}
\usepackage{graphics}

\usepackage[dvips]{graphicx}
\usepackage{amsmath}

\begin{document}

%
%

\title{From single-particle physical distributions to probabilistic measures of two-particle entanglement}

%
%
%
\author{I. Nagy}
\affiliation{Department of Theoretical Physics,
Institute of Physics, \\
Budapest University of Technology and Economics, \\ H-1521 Budapest, Hungary}
\affiliation{Donostia International Physics Center, P. Manuel de
Lardizabal 4, \\ E-20018 San Sebasti\'an, Spain}
%
%
\author{M. L. Glasser}
\affiliation{Department of Physics, Clarkson University, Potsdam,\\
New York 13699-5820, USA}
\affiliation{Departamento de F\'{i}sica Te\'orica, At\'omica y \'Optica, Universidad de Valladolid, \\
E-47071 Valladolid, Spain}
\affiliation{Donostia International Physics Center, P. Manuel de
Lardizabal 4, \\ E-20018 San Sebasti\'an, Spain}

\date{\today}
\begin{abstract}

An inversion method is formulated  for extracting entanglement-related
information on two-particle interactions in a one-dimensional system from measurable one-particle 
position- and momentum-distribution functions. The method is based on a
shell-like expansion of these norm-1 measured quantities in terms of product states taken from
a parametric orthonormal complete set. The mathematical constraints deduced from  these point-wise expansions are
restricted by the underlying physics of our harmonically confined and interacting  Heisenberg model.
Based on these exact results, we introduce an approximate optimization scheme for  different inter-particle interactions
 and discuss it from the point of view of entropic correlation measures.

\end{abstract}


\maketitle

\section{Motivation}

Let us take the Hamiltonian, in atomic units, of a {\it single} harmonic oscillator
\begin{equation}
\hat{H}_0(x)\, =\, -\, \frac{1}{2} \frac{d^2}{dx^2}\, +
\frac{1}{2}\, \omega_0^2{x}^2,
\end{equation}
the ground-state normalized solution  to the Schr\"odinger equation is $\phi_0(x,\omega_0)=(\omega_0/\pi)^{1/4}\exp(-\omega_0 x^2/2)$. 
Its square gives the normalized density distribution function, $n(x)=[\phi_0(x)]^2$, whose 
Fourier transform is
\begin{equation}
n(k)=\frac{1}{\sqrt{2\pi}}\, e^{-k^2/(4\omega_0)}. 
\end{equation}
This function can be sampled by X-ray scattering, and by Fourier inversion, one may say that $n(x)$ is 
accessible experimentally. The normalized one-particle momentum density distribution, $f(k)$, 
is, on the other hand,  connected to Compton scattering. 
The relation between these position and momentum-space
distributions, i.e., between the two sets of experimental data, is one-to-one 
because $\phi_0(x)=\sqrt{n(x)}$ and $f(k)$ is given by
\begin{equation}
f(k)\, =\, [\phi_0(k,\omega_0)]^2\, =\, \left[\left(\frac{1}{\pi\omega_0}\right)^{1/4}\,
e^{-k^2/(2\omega_0)}\right]^2. 
\end{equation}

From $\phi_0(x,\omega)$ and $\phi_0(k,\Omega)$ we can determine the corresponding kinetic
energies
\begin{equation}
<K_1>\, =\, \frac{1}{2}\, \int_{-\infty}^{\infty} dx \left|\frac{d\phi_o(x,\omega)}{dx}\right|^2
= \frac{1}{4}\, \omega
\end{equation}
\begin{equation}
<K_2>\, =\, \int_{-\infty}^{\infty}\frac{1}{2}\, k^2\, f(k,\Omega)\, dk = \frac{1}{4}\, \Omega
\end{equation}
In the noninteracting case, $\omega=\omega_0$ and $\Omega=\omega_0$, these are equal, but  in the interacting case  below, ($\omega_0\Rightarrow{\Omega_s}$) 
only  Eq.(5) gives the exact result. 
The definition in Eq.(4), even with the exact density ($\omega_0\Rightarrow{\omega_s}$)
\cite{Dreizler90} gives $[<K_1>/<K_2>]<1$.

When the Gaussian function $f(k)$ is characterized by $\Omega_s\neq{\omega_0} $, instead of 
$\omega_s$ used for the density distribution, the one-to-one mapping is lost. In case of  
harmonic external confinement, one should conclude from this observation  that
we have at least two similar particles in the ground-state under confinement and, moreover, 
these particles are dynamically correlated.
Is there, in the physically most important interacting case, still something 
useful which one can extract from the above two observables? 
How  could  a  proper mathematical recipe be formulated? These are the 
questions motivating this note on the application of {\it reduced} information. 
We feel that  examining the weakly interacting two-particle case
is a conservative first step having general information-theoretic relevance.

\newpage

\section{Decompositions, constraints, and optimization}

Four our interacting system, there is no simple connection between the normalized position-space
density $n(x)$ and the normalized momentum-space density $f(k)$, despite the fact that they are  related by the Fourier transformation between the wave functions, and associated density matrices, in these spaces \cite{Dreizler90}. Furthermore, the reduced one-particle density matrices, 
the sources of $N(x)$ and $F(k)$ are two-variable functions. They
contain more information than the corresponding one-variable probability densities, i.e., their
diagonals. Besides, since  correlation is encoded differently \cite{Harriman91}
in the observables discussed above,  a successful recipe for extracting information must rest on both  
densities. Such {\it chameleon}-like behavior in the observables requires care in their mathematical treatment. 
One can not simply follow the Duke of Gloucester who, in  Shakespeare's famous play Henry VI, 
stated: "I can add colors to the chameleon".

The details of the one-to-one correspondence outlined above for the one-particle 
case suggest that decompositions of  $N(x)$ and $F(k)$, when properly normalized, 
into {\it products} of one-variable functions belonging to complete sets, 
and the Fourier-transformation $(\mathcal{F})$ link 
between these sets, could form the mathematical basis of the recipe. 
In  the extraction of information, one requires spatially-independent  decomposition weights. 
Since these weights are eigenvalues of the underlying one-body matrices
they allow a detailed analysis of  entanglement entropies. However,
the two-variable eigenfunctions, i.e., the natural orbitals, 
are not directly accessible experimentally. Thus,
physically important information, say the energy scales behind extensions of these
optimal orbitals in real space, remains intact. 

In this note we consider a {\it weakly} interacting two-particle system under 
 common harmonic confinement $(1/2)\omega_0^2(x_1^2 + x_2^2)$. This 
is the standard condition, e.g., in recent experiments on optically trapped systems with controllable
number of constituents \cite{Wenz13}.
For this system we introduce, assuming $\omega_s\neq{\Omega_s}$, the two Gaussians 
\begin{equation}
\phi_1(x,\omega_s)\, =\, \left(\frac{\omega_s}{\pi}\right)^{1/4} e^{-\frac{1}{2}\omega_s x^2} 
\end{equation}
\begin{equation}
\phi_2(k,\Omega_s)\, =\, \left(\frac{1}{\pi\Omega_s}\right)^{1/4} e^{-\frac{1}{2}\frac{k^2}{\Omega_s}}.
\end{equation}
in order to {\it model} the correlated density distribution function, $N(x)=[\phi_1(x)]^2$, and the
correlated momentum distribution function, $F(k)=[\phi_2(k)]^2$.  
The   case $\omega_0=\omega_s=\Omega_s$ obviously  corresponds to the 
noninteracting situation where, of course, $\phi_2(k)=\mathcal{F}[\phi_1(x)]$.

\newpage

The desired product-representations of the experimental data-functions, $N(x)=[\phi_1(x)]^2$ and $F(k)=[\phi_2(k)]^2$, with common spatially-independent weighting coefficients necessary for 
a linear mapping, such as Fourier transformation, are given by 
\begin{equation}
N(x,\omega_s)\, =\, \sum_{m=0}^{\infty}\, (1-Z)Z^m\,
[\phi_m(\sqrt{\bar{\omega}}x)]^2
\end{equation}
\begin{equation}
F(k,\Omega_s)\, =\, \sum_{m=0}^{\infty}\, (1-Z)Z^m\, 
[\phi_m(k/\sqrt{\bar{\omega}})]^2
\end{equation}
\begin{equation}
\phi_m(\sqrt{\alpha}u)\, = \,
\left(\frac{\alpha}{\pi}\right)^{1/4}\frac{1}{\sqrt{2^m\,
m!}}\, e^{-\frac{1}{2}{\alpha}\, u^2}\,
H_m(\sqrt{\alpha}u).
\end{equation}
Here $\bar{\omega}$ is an orbit-parameter. For Gaussian densities these Mehler's \cite{Erdelyi53}
representations are point-wise \cite{Riesz55}, and $\sum_{m=0}^{\infty}(1-Z)Z^m=1$. 
The constraints on the expansions are
\begin{equation}
\omega_s\, =\, \bar{\omega}\, \frac{1-Z}{1+Z}
\end{equation}
\begin{equation}
\frac{1}{\Omega_s}\, =\, \frac{1}{\bar{\omega}}\, \frac{1-Z}{1+Z},
\end{equation}
from which $\bar{\omega}=\sqrt{\omega_s\Omega_s}$ and $Z=[1-\sqrt{\omega_s/\Omega_s}]/
[1+\sqrt{\omega_s/\Omega_s}]\leq{1}$. Since we know from the physics \cite{Dreizler90} of kinetic 
energy that $[<K_2(\Omega_s)>/<K_1(\omega_s)>]\geq{1}$, we have $(\Omega_s/\omega_s)\geq{1}$. 

A useful probabilistic measure of correlation is the purity $\Pi$. From the properties
of our normalized occupation numbers, $P_m\equiv{(1-Z)Z^m}$, this measure is given by
\begin{equation}
\Pi\, =\, \sum_{m=0}^{\infty}(P_m)^2\, =\, \frac{1-Z}{1+Z}\, =\, \sqrt{\frac{\omega_s}{\Omega_s}}\, \leq{1},
\end{equation}
in terms of the ratio of $\omega_s$ and $\Omega_s$, which characterize
the experimentally accessible distributions in position and momentum spaces, respectively.
Related, commonly applied information-theoretic quantities are \cite{Renyi70}  R\'enyi's ($R$) and von
Neumann's ($N$) entropies
\begin{equation}
S_R(q)\, =\, \frac{1}{1-q}\,
\ln\frac{(1-Z)^q}{1-Z^q} 
\end{equation}
\begin{equation}
S_N = -\left[q^2 \frac{d}{dq} \left(\frac{1-q}{q} S_R(q)\right)\right]_{q=1}\,
=\, -\ln(1-Z)\, -\frac{Z}{1-Z} \ln{Z}. 
\end{equation}
Von Neumann's $S_N$ is the entropy of thermodynamics. 
But, in agreement with an earlier remark \cite{Srednicki93}, the above measures depend 
solely on a ratio of physical parameters.  
For entropies, taken at arbitrary $q$ values,
the orbit-extension parameter, $\bar{\omega}$, is not needed. 
Therefore, pure information-theoretic measures alone are not applicable directly
to determine scale-dependent physical quantities. 
Determining the {\it sign} of the inter-particle interaction (see, below) could be a nontrivial problem 
for reverse engineering, due to duality \cite{Pipek09,Glasser13,Schilling14}.

\newpage

To proceed in our realistic modeling of  a confined system \cite{Wenz13}, we add to the
Schr\"odinger Hamiltonian  the tunable (via $\lambda$, see below) two-particle interaction  
\begin{equation}
\hat{H}(x_1,x_2)\, =\, -\, \frac{1}{2}\left(\frac{d^2}{dx_1^2}\, +
\frac{d^2}{dx_2^2}\right) +\frac{1}{2}\, \omega_0^2({x}_1^2+{x}_2^2) +
v_{H}(|x_1-x_2|),
\end{equation}
When $v_H\neq{0}$, we get $\Omega_s\neq{\omega_0}$, $\omega_s\neq{\omega_0}$, 
$\Omega_s\neq{\omega_s}$.  In order to have a rigorous foundation for understanding
 information-extraction from observables, we turn to the 
specific \cite{Heisenberg26} interaction $v_{H}(|x_1-x_2|=\lambda (\omega_0^2/2)(x_1-x_2)^2$. 
Based on this interaction, we recently  derived closed-form expressions \cite{Nagy17} 
for both two-variable one-matrices 
\begin{equation}
\Gamma_1(x_1,x_2)=\phi_s(x_1)\, \phi_s(x_2) 
\times{e^{-D[(x_1-x_2)/\sqrt{2}]^2}}
\end{equation}
\begin{equation}
\Gamma_1(k_1,k_2)=\frac{1}{\sqrt{\pi(\omega_s+2D)}}\, e^{-\frac{1}{2}(k_1^2+k_2^2)\frac{\omega_s+D}
{\omega_s(\omega_s+2D)}}\, e^{+\frac{Dk_1k_2}{\omega_s(\omega_s+2D)}} 
\end{equation}
where, with $\omega_s\equiv{2\omega_1\omega_2/(\omega_1+\omega_2})$, we introduced 
the following abbreviations
\begin{equation}
\phi_s(x)=\left[\frac{\omega_s}{\pi}\right]^{1/4}\,
e^{-\frac{1}{2}\omega_s\, x^2}     
\end{equation}
\begin{equation}
D\, =\, \frac{1}{4}\, \frac{(\omega_1-\omega_2)^2}
{\omega_1+\omega_2}\, \geq{0}. 
\end{equation}
We derive $\omega_1=\omega_0$ and $\omega_2=\omega_0\sqrt{1+2\lambda}$ in the underlying \cite{Nagy17} 
normal-mode separation of the
Schr\"odinger equation with Eq.(16). Thus, for the repulsive harmonic inter-particle
interaction, the allowed range is $\lambda\in{(-0.5,0]}$. Both $\omega_s$ and $D$, and therefore $Z$,
show a dual character \cite{Pipek09,Glasser13,Schilling14}. This means that to any allowed repulsive coupling there exists a corresponding attractive one with the same value for $Z$.
Clearly, with Heisenberg's  
inter-particle interaction model \cite{Heisenberg26} the measurable 
$N(x)=\Gamma_1(x,x)$ and $F(k)=\Gamma_1(k,k)$ 
are known theoretically. 
Therefore, in this case,
we get $\Omega_s=\omega_s+2D=\omega_1+\omega_2$, by which the kinetic energy becomes $<K_2>=(1/4)(\omega_s+2D)>(1/4)\omega_s=<K_1>$. 
But, and this is crucial to information-extraction, by taking the diagonal of Eq.(17) 
to get a measurable distribution 
we have no {\it separate} access to $D(\omega_1,\omega_2)$ which produces non-idempotency.

Finally, we turn to an optimization procedure which may connect two Schr\"odinger Hamiltonians. 
One could replace, following a recent proposal \cite{Brabec13}, a realistic
two-particle Hamiltonian having non-harmonic inter-particle interaction
by the Heisenberg ($H$) Hamiltonian. For instance, one could apply  total-energy
correspondence as  a constraint on such replacement. One may argue, of 
course, that the $\lambda$-coupling in Eq.(16)
has been chosen  qualitatively and it is this mapping correspondence which would allow it to be
determined quantitatively. We expect, based on physical considerations, 
that such an optimized correspondence between two Hamiltonians can be reasonable only if the harmonically confined \cite{Wenz13} particles interact {\it weakly}. 
At that small coupling the non-idempotency driver
scales as $D\sim{\lambda^2}$, i.e., the deviations of $\omega_s$ and $\Omega_s>\omega_s$ 
from $\omega_0$ are small. The associated entropies are small as well.
However, at stronger couplings, one may get a 
serious problem by applying such an optimization scheme to information-theoretic measures. 

We now quantify this problem by considering the attractive ($\lambda>0$) case in Eq. (16). 
Say we construct a more  realistic model by taking for the inter-particle interaction
\begin{equation}
v_{C}(|x_1-x_2|)\, =\, \frac{\Lambda}{(x_1-x_2)^2} \nonumber
\end{equation}
with $\Lambda>0$. We focus here on the  strong coupling limit \cite{Koscik15,Anna15}.
 From the prescribed equivalence of ground-state
energies $E_H(\lambda\rightarrow{\infty})\propto{\sqrt{1+2\lambda}}$ 
and $E_C(\Lambda\rightarrow{\infty})\propto{\sqrt{1+4\Lambda}}$, 
with $v_H(\lambda)$ and $v_C(\Lambda)$ respectively, we get
the simple  correspondence $\lambda=2\Lambda$. However,
with the  singular interaction above. one gets, in the associated Wigner-crystal limit at 
strong coupling, $\Lambda$-independent occupation numbers \cite{Koscik15,Anna15}. 
Thus the purity is approximately  $\Pi_C(\Lambda\rightarrow{\infty})\simeq{0.528}$. 
In the energetically optimized Heisenberg case, i.e., 
with a harmonic  interaction $v_H(\lambda=2\Lambda)$, we obtain the different 
behavior, $\Pi_H(\Lambda)\simeq{\sqrt{2}/\Lambda^{1/4}}$ at $\Lambda\rightarrow{\infty}$. 

This, seemingly,  moderate numerical difference in 
an information-theoretic measure is related,  physically, to crucially different behaviors
of the underlying wave function as a function of the relative coordinate. Our quantitative
observation at strong coupling is not in contradiction with the prediction 
 \cite{Brabec13} which relies on perturbation theory.
Clearly, one can only get a physically reasonable approximation for the linear entropy, $L_C=1-\Pi_C$, 
 at small (i.e., perturbative) coupling within the proposed optimization framework.

\newpage

\section{Summary}

An inversion method is formulated for extracting entanglement-related
information on two-particle interactions from measurable one-particle
distribution functions in position and momentum spaces. The method is based on
shell-like expansions of these measurable norm-1 quantities  in terms of properly 
weighted product states taken from a parametric  complete orthonormal  set. 
It is found that without further physical details, encoded in the two-variable
reduced one-particle density matrices, an unambiguous characterization of the
inter-particle interaction is not possible by inverting such information.

We have, therefore, given a concrete answer to Pauli's general question \cite{Ballentine98} 
of whether the position and momentum probability densities are sufficient to determine the 
statistical state operator. These distributions are not sufficient. One method for resolving 
the dual character in the sign of an inter-particle interaction is to make
use of the dynamical evolution \cite{Nagy17} of the correlated state. In such evolution, one of 
the normal-mode frequencies, $\omega_2=\omega_0\sqrt{1+2\lambda}$ in the Heisenberg model, 
could be measurable via the corresponding breathing mode \cite{Brabec13}.

Based on exact results obtained with Heisenberg's Hamiltonian, a recently suggested 
optimization procedure for introducing a different inter-particle interaction
is formulated and analyzed quantitatively from the point of view of entropic correlation measures.
This analysis shows that, as expected on physical grounds, an energy-based optimization 
scheme could be useful only at weak inter-particle couplings.


%
\begin{acknowledgments}
The authors thank Professor Ricardo D\'{i}ez Muino for the 
warm hospitality at the DIPC. The kind help of Professor Anna Okopinska is gratefully acknowledged. MLG  acknowledges  the financial support of MINECO (Project MTM2014-57129-C2-1-P) 
and Junta de Castilla y  Leon (UIC 011).
\end{acknowledgments}

\end{document}